\documentclass[aps,pra,preprint,superscriptaddress,amsmath,amsfonts,showpacs]{revtex4}
\usepackage{amsmath, amsthm, amssymb}
\usepackage{graphics}
\usepackage{graphicx}
\usepackage{epstopdf}
\usepackage{ulem}

\begin{document}

\title{Quantum Control of Photodissociation via Manipulation of Bond Softening}

\author{Adi Natan} \thanks{Current Address: PULSE Institute, SLAC National Accelerator Laboratory, and Department of Applied Physics, Stanford University, 2575 Sand Hill Rd, Menlo Park, CA 94025, USA}

\affiliation{Department of Physics of Complex Systems, Weizmann Institute of Science,
Rehovot 76100, Israel}

\author{Uri Lev}
\affiliation{Department of Particle Physics and Astrophysics, Weizmann Institute of
Science, Rehovot 76100, Israel}

\author{Vaibhav S. Prabhudesai} \thanks{Current Address: Dept. of Nuclear and Atomic Physics, Tata Institute of Fundamental Research, Homi Bhabha Road, Colaba, Mumbai 400005 India}
\affiliation{Department of Physics of Complex Systems, Weizmann Institute of Science,
Rehovot 76100, Israel}

\author{Barry D. Bruner}
\affiliation{Department of Physics of Complex Systems, Weizmann Institute of Science,
Rehovot 76100, Israel}

\author{Daniel Strasser}
\affiliation{Institute of Chemistry, Hebrew University, Jerusalem 91904, Israel}

\author{Dirk Schwalm}
\affiliation{Department of Particle Physics and Astrophysics, Weizmann Institute of
Science, Rehovot 76100, Israel}
\affiliation{Max-Planck-Institut f\"{u}r Kernphysik, Heidelberg 69117, Germany}

\author{Itzik Ben-Itzhak}
\affiliation{J. R. Macdonald Laboratory, Department of Physics, Kansas State
University, Manhattan, Kansas 66506, USA}

\author{Oded Heber}
\affiliation{Department of Particle Physics and Astrophysics, Weizmann Institute of
Science, Rehovot 76100, Israel}

\author{Daniel Zajfman}
\affiliation{Department of Particle Physics and Astrophysics, Weizmann Institute of
Science, Rehovot 76100, Israel}

\author{Yaron Silberberg}
\affiliation{Department of Physics of Complex Systems, Weizmann Institute of Science,
Rehovot 76100, Israel}

\begin{abstract}
We present a method to control photodissociation by manipulating the bond softening
mechanism occurring in strong shaped laser fields, namely by varying the chirp sign and
magnitude of an ultra-short laser pulse. Manipulation of bond-softening is
experimentally demonstrated for strong field ($10^{12} - 10^{13} $ W/cm$^2$)
photodissociation of H$_2^+$, exhibiting a substantial increase of dissociation by
positively chirped pulses with respect to both negatively chirped and transform
limited pulses. The measured kinetic energy release and angular distributions are
used to quantify the degree of dissociation control. The control mechanism is attributed to the interplay of dynamic alignment and chirped light induced potential curves.
\end{abstract}


\maketitle

\section{Introduction}

Controlling chemical processes using laser pulses has been a longstanding
goal of scientists in physics and chemistry.  Much effort has been directed in recent
years towards achieving efficient photodissociation via quantum coherent control. One
commonly used approach is based on self-learning techniques, in which an optimal
pulse shape is found through various iterative optimization procedures
\cite{JudsonAssion,Assion,Weinacht}. However, while learning algorithms have been successful in
producing effective control fields, the physical mechanisms underlying this control
are still generally difficult to interpret. Another category of control schemes
involves adiabatic population transfer methods, where chirped pulses are used in order
to efficiently populate an excited level \cite{Chelkowski,Shore}, or to control
 dissociative ionization \cite{Frasinski,Breunig,Singh}. Yet a different
approach is to employ a pump-probe scheme, where a typical time scale of the system,
such as the vibrational period, is used in controlling the outcome
\cite{Niikura,Niikura2,Sussman,Ergler,Kelkensberg}. Here, we demonstrate how
light-matter interaction can be harnessed to achieve control using the bond softening
mechanism.

\section{Theoretical Background}

\begin{figure}
\begin{center}
\includegraphics[width=1\columnwidth]{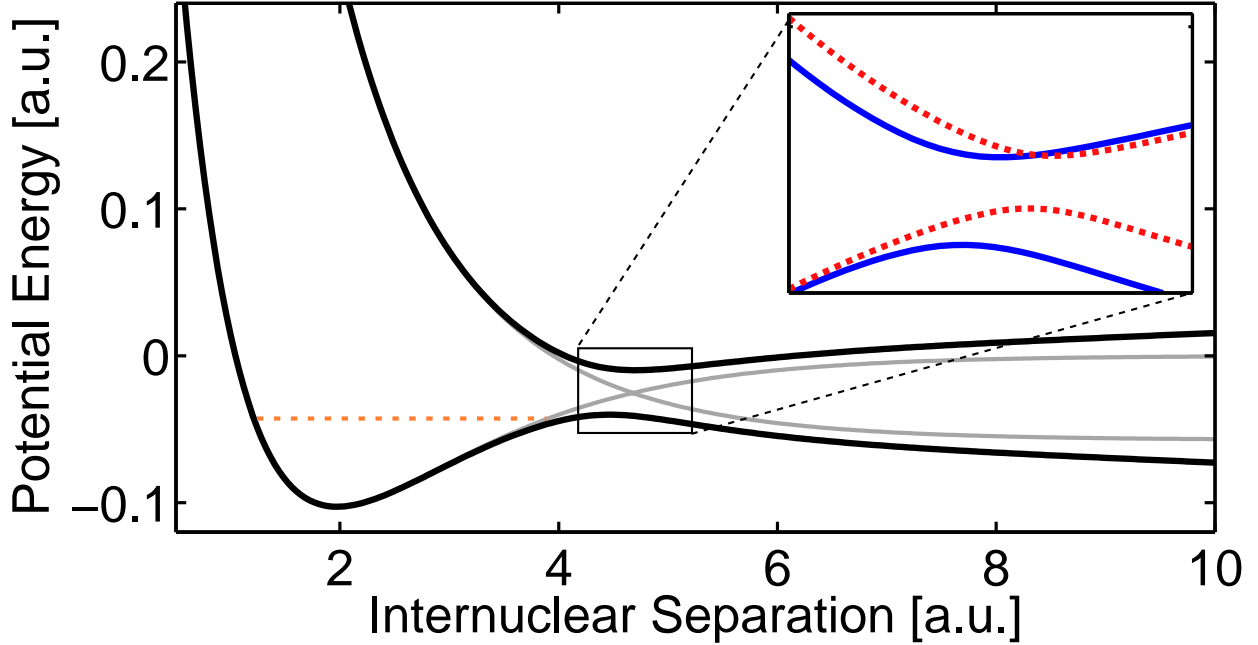}
\end{center}
\caption{(color online)  H$_2^+$ potential curves. As intensity increases,
the adiabatic (black) light induced potential curves are modified and an avoided
crossing opens at the resonant internuclear separation, causing dissociation for
levels such as $v=7$ (dotted orange)
. The dressed diabatic potential curves (grey) show the position
of the crossing. (inset) For positively chirped pulses, the adiabatic
potential curves are dynamically modified from red (dashed red) to blue (solid blue)
shifted potential curves, according to the direction of the frequency sweep.
A larger gap size for blue-detuned frequencies develops as the temporal alignment
evolves close to peak intensity.} \label{potential_curves}
\end{figure}

 The interaction of molecules with  strong fields can be represented by the Floquet dressed state formalism \cite{Giusti-Suzor}. 
   For the case of diatomic molecules with bound and repulsive potential curves, like H$_2^+$, the interaction of light with the molecule shifts the energy of the upper, repulsive potential curve down by $\hbar \omega$, where $\omega$ is the frequency of the laser.  The new diabatic potential curves now cross at an internuclear distance where the two electronic states are resonantly coupled  by the laser field (Fig. \ref{potential_curves}). As the laser intensity increases an avoided crossing opens at the resonant internuclear separation, and in addition, the molecule starts to dynamically align with the field polarization. Such excitation fields can then be sufficient to cause bound population to cross the
barrier and dissociate, in a process known as bond softening \cite{Bucksbaum90}. This adiabatic picture can still be used when describing non-adiabatic behavior,  as long as the pulse duration is not shorter than the molecular vibration time-scale \cite{posthumus}.

 Previous studies that
used such phenomena to control photodissociation employed a pump-probe scheme. This
technique often required a precursor molecule in order to launch a vibrational wave
packet on the ground state of the target molecular ion. The control of
photodissociation depended on the precise timing of the wave-packet motion to the crossing
point where the gap opens. This timing was usually controlled by varying the time
delay between a strong pump pulse that excites a wave-packet, and a `control' pulse
that is responsible for creating the avoided crossing
\cite{Niikura,Niikura2,Sussman}. In contrast to this approach, we show that it is
possible to control strong field photodissociation by manipulation of the avoided
crossing using intense shaped pulses, without the need of an initial wave-packet
preparation.

We focus on one of the most studied molecular systems, H$_2^+$. This simple system
has revealed remarkably rich dynamics such as bond hardening, bond softening and
above threshold dissociation
\cite{Wang,sandig,Giusti-Suzor,posthumus,Bucksbaum90,McKenna}. For H$_2^+$ only the
two lowest-lying potential surfaces, $1s \sigma_g$, and $2p \sigma_u$, need to be
considered for the field intensities used in the present experiment. Bond softening can be
described by diagonalizing the molecule-field Hamiltonian in the Floquet picture \cite{Giusti-Suzor} (see Fig. \ref{potential_curves}):
\begin{equation}
 E_{\pm} =  \frac{V_g+V_u-\hbar \omega}{2} \pm \frac{1}{2}
\sqrt{(V_g+\hbar \omega -V_u)^2+4 V_{gu}^2}
\end{equation}

\noindent where  $V_g(R)$ and $V_u(R)$ are the molecular potential energies of the $1s
\sigma_g$ and $2p \sigma_u$ states respectively, $V_{gu}(R) =-\frac{1}{2} \varepsilon(t) D(R) \cos\theta(t) $ is
the coupling term between them with the transition dipole moment $D(R)$,  $\theta(t)$
is the angle of the molecular axis with respect to the laser polarization, and
$\varepsilon(t)= A(t) \cos[\omega(t)t]$, where $A(t)$ is the temporal field envelope
with instantaneous frequency $\omega(t)$. The properties of the avoided crossing are thus governed by the field intensity and frequency, and the degree of molecular alignment with respect to the laser polarization. The temporal evolution of molecular alignment results in a change of the angular distribution of the molecules and is dependent on the intensity of the excitation pulse \cite{Uhlmann}.

Dynamically modifying the avoided crossing can therefore be used to
control dissociation. The position of the avoided crossing depends on $\hbar \omega(t)$.
For example, positively chirped pulses will displace the avoided
crossing according to the direction of the frequency sweep, from red to blue-detuned
frequencies. The chirp rate corresponds to the displacement rate of the avoided
crossing, the longer the pulse the slower the frequency sweep. The gap at the
avoided crossing is proportional to $V_{gu}$ and becomes larger for increasing $A(t)$ and $\cos\theta(t)$. In addition,
$\cos\theta(t)$ significantly increases as the field amplitude peaks, but only slightly decreases afterwards \cite{Uhlmann,esry2009}. Thus, for typical symmetric temporal
envelopes such as gaussian pulses, the energy gap at the avoided crossing will be
larger at $t_0+\Delta t$ than at $t_0-\Delta t$, where $t_0$ the time of peak
amplitude, and $\Delta t$ is some positive time delay. Specifically for positively
chirped pulses, the avoided crossing around the temporal peak of the pulse will be larger for
blue-detuned frequencies than for red-detuned ones, promoting dissociation of
levels further below the crossing (Fig. \ref{potential_curves} inset).  For negatively chirped
pulses however, the dynamic displacement raises the effective barrier and
suppresses dissociation of these levels.


\section{Experimental}

We examined photodissociation of H$_2^+$ experimentally by
focusing 30 fs pulses at 795 nm central wavelength (peak intensities
$\leq2\times10^{13}$ W/cm$^2$) on a well collimated, pulsed, 4 keV H$_2^+$ ion beam.
The molecules were in Franck-Condon distribution of
vibrational states. Both neutral H and proton fragments of the molecular ion were measured in coincidence using a time and position sensitive detector comprised of a phosphor screen anode attached to a microchannel plate with a CCD camera and a home-built frame grabber. The full 3D momentum components for each event were reconstructed from the measured positions and times of flight. From these momentum vectors, we
deduced the kinetic energy release (KER) upon dissociation and the angle $\theta$
between the molecular axis and the laser polarization at the time of dissociation.
Post-dissociation rotations were neglected as the experimental parameters were within
the axial recoil approximation \cite{esry2009}. The pulse shaping was performed using
a conventional 4-f phase-only pulse shaper \cite{Weiner}. Further details regarding
the experimental system have been described elsewhere \cite{shifts}. Using the
pulse shaper, we applied a quadratic spectral phase function, $\Phi (\omega)=
\phi^{''}(\omega-\omega_0)^2/2$, where $\phi^{''}$ is the group dispersion delay
(GDD) parameter, which corresponds to the chirp magnitude, and $\omega_0$ is the
central laser frequency. The pulse shaper was also used to accurately compensate
higher orders of dispersion that can arise from downstream optical elements. During
the experiments the sign and magnitude of the chirp were alternated every three
minutes in order to avoid possible long term drift biases in the measurements.

\begin{figure}
\begin{center}
\includegraphics[width=1 \columnwidth]{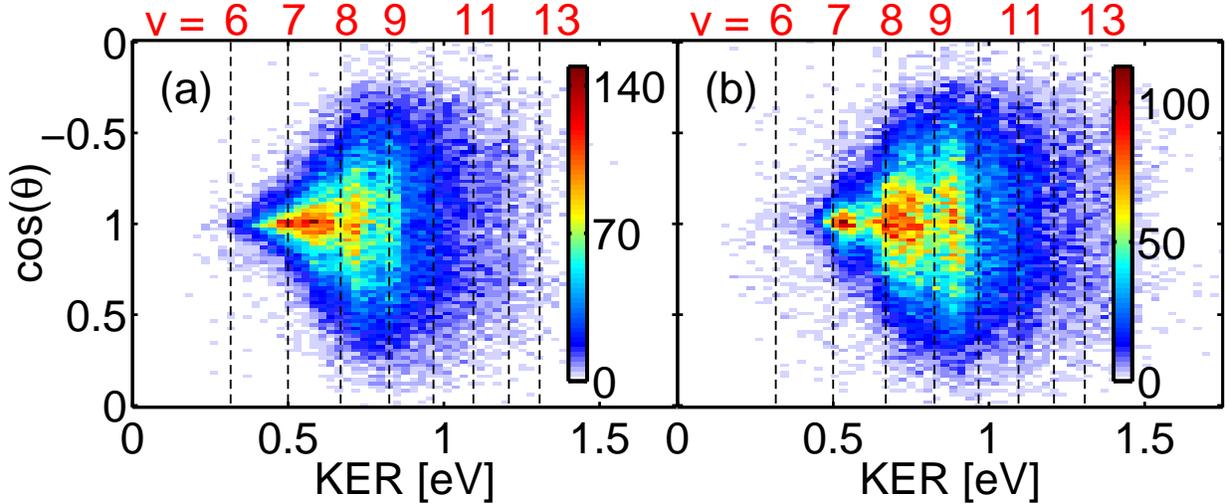}
\end{center}
\caption{(color online) Dissociation yield of H$_2^+$ as a function of KER and $\cos\theta$ by (a) positively chirped  and (b) negatively chirped
pulses of 90 fs duration (GDD= $\pm$ 920  fs$^2$, peak intensity of \textit{I$_0$} =
$7 \times 10^{12}$ W/cm$^2$).  Vertical dotted lines indicate the expected peak
positions of different field-free vibrational levels at the central wavelength of the laser
(795nm).} \label{ker90fs}
\end{figure}

\section{Results and Discussion}

 We measured the KER spectra for pulses with GDD values ranging from -1920 to +1920 fs$^2$. This
resulted in negatively and positively chirped pulses of durations of 30 fs (transform
limited) to 180 fs, of the same bandwidth and fluence. Fig. \ref{ker90fs}(a-b) show
density plots of the distribution of dissociation events as a function of KER and
$\cos\theta$ for 90 fs (GDD of $\pm 920$ fs$^2$) positively and negatively
chirped laser pulses, respectively. There are key differences between the KER spectra
for the two chirp signs. First, we resolve energetic shifts of KER peaks that result
from the temporal ordering of frequencies in the chirped pulse \cite{shifts}.
Second, we note significant changes in the angular distributions and signal strengths
in the KER region below $0.7$ eV associated with dissociation of vibrational levels below the $1s
\sigma_g - 2p \sigma_u$ one-photon curve crossing. These changes reflect the different character of
the dissociation mechanism for the field intensity used. Vibrational levels that are
located close to the one-photon crossing, as in the case of $v=9$, are expected to
dissociate at the leading edge of the pulse \cite{shifts}, with angular distributions
of $\cos^2\theta$. This distribution indicates that a resonant one-photon transition
occurred and that no alignment took place \cite{sandig,williams}. However,
dissociation from lower levels, such as $v=7$, show narrower angular distributions
that can be expressed by higher cosine powers \cite{sandig,Wang2005}, reflecting
geometric or dynamic alignment \cite{posthumus}. We label such non-resonant dissociation events as
`Over the Barrier Dissociation' (OBD).

\begin{figure}
\begin{center}
\includegraphics[width=1 \columnwidth]{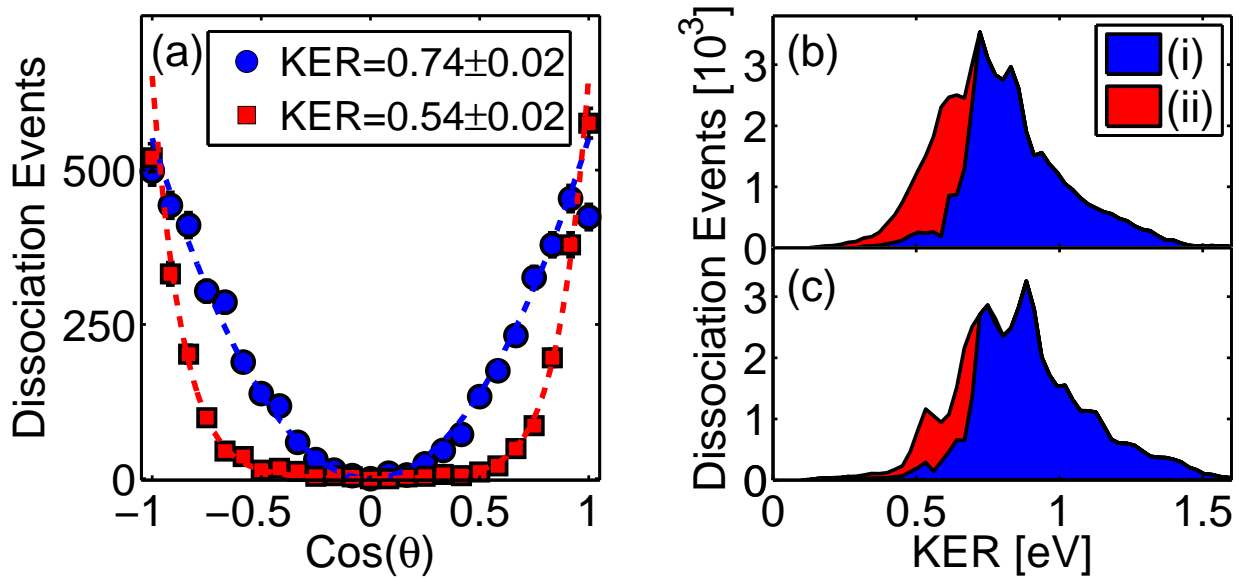}
\end{center}
\caption{(color online) Analyzing H$_2^+$ dissociation angular distributions. (a) Angle vs
dissociation events for specific KER values for positively chirped pulses
(GDD=920 fs$^2$) and the corresponding fits (dashed lines). For KER=0.54 eV a $\cos^8\theta$
distribution fits suggesting OBD, whereas for  KER=0.74 eV a $\cos^2\theta$
distribution fits suggesting near-resonant dissociation. (b) The dependence of the $W_{n\geq2}$ contribution (red area) and $W_1$ contribution (blue area) on the KER value, for GDD=+920 fs$^2$ and (c)
GDD=-920 fs$^2$ (see text).}

 \label{ang_dist_fit}
\end{figure}

We used the angular distribution information to separate the OBD contribution from
the near-resonant dissociation events. This approach was taken to avoid the KER
shifts that arise in near-resonant dissociation events for different chirp
magnitudes \cite{shifts}. The vibrational level that is nearest to resonance with the
laser is $v=9$, whereas $v=7$ has no overlap with the laser bandwidth, hence can
dissociate only via OBD. Therefore, we can use these two levels to demonstrate the controllability of OBD and near-resonant dissociation.
For example, in Fig. \ref{ang_dist_fit}(a) we present two angular distributions due
to a 90 fs positively chirped pulse of GDD=920 fs$^2$ that relate to levels $v=9$ and
$v=7$, for which KER=0.74$\pm$0.02 and 0.54$\pm$0.02 eV, respectively. A cosine power
series function of the form: $f(E,\theta)=\sum_{n} a_n(E) \cos^{2n}\theta$ was fitted
to the measured angular distributions for the given KER values.
    Having the fit factors $a_n(E)$, we calculated the relative contribution of the n$^{th}$ term
 for a given KER by  $W_n(E) = \int  \frac{a_n(E) \cos^{2n}\theta}{f(E,\theta)}d
\cos\theta $. We found that $n=4$ is the minimal power required to obtain an accurate
fit for the lower KER range (i.e. around 0.54eV). Near-resonant dissociation contribution will therefore be expressed by $W_1$,
and OBD contribution by $W_{n\geq2}$.

\begin{figure}
\begin{center}
\includegraphics[width=1\columnwidth]{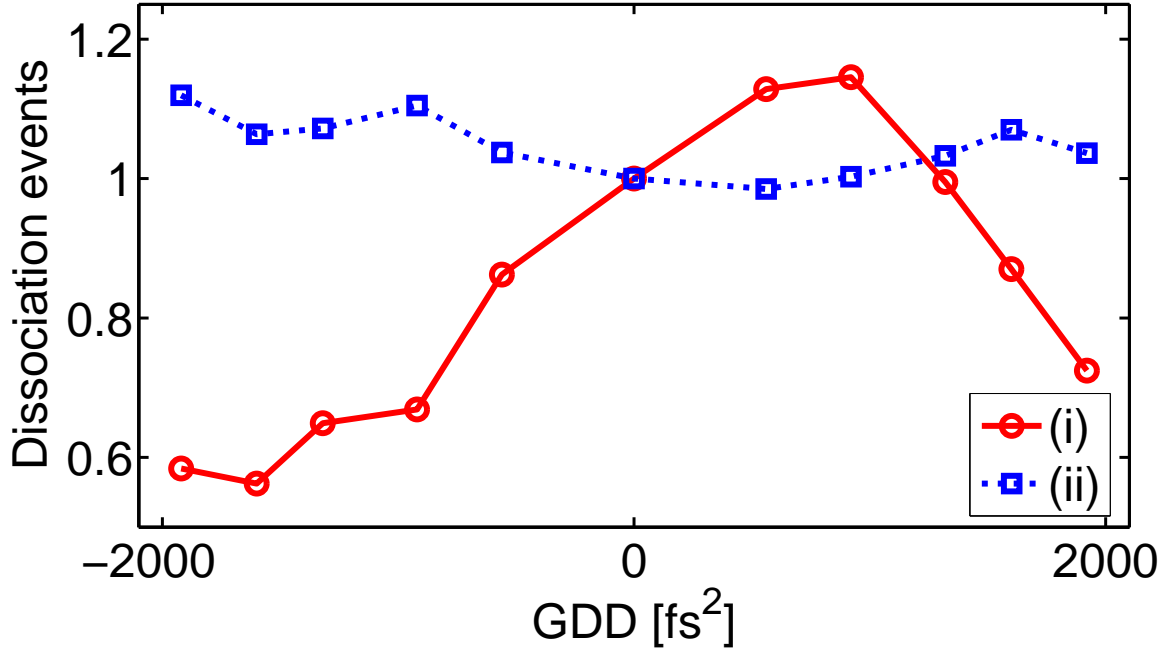}
\end{center}
\caption{(color online) Controlling OBD dissociation events. (i) For OBD, positive chirp is more efficient
than negative chirp, whereas (ii) near-resonant dissociation is less affected
by different chirp magnitudes. Both curves are normalized to 1 at zero chirp.} \label{fig_4}
\end{figure}

The fitting procedure  was used to analyze the angular distribution of the entire KER spectrum for each chirp magnitudes up to $\pm 1920$ fs$^2$. For example, in Fig. \ref{ang_dist_fit}(b-c) we project all dissociation angles on the KER axis for
positively and negatively chirped pulses of GDD $=\pm 920$ fs$^2$. The $W_1$
contribution (blue areas) is similar for the two chirp signs, with about 10$\%$
difference. However, the $W_{n\geq2}$ (red areas) contribution shows $70\%$ more OBD for
+920 fs$^2$ than for -920 fs$^2$. The overall number of near-resonant and OBD events
 was then extracted for different chirp magnitudes and shown in Fig. \ref{fig_4}.
 For near-resonant dissociation we obtain less than 10$\%$ variation in dissociation
 rate, although the peak intensity changes
 by a factor of six for different chirp magnitudes. This finding supports the assumption that the
  fragments measured with $\cos^2\theta$ angular distributions ($W_1$) were
  dissociated mostly by resonant one-photon transitions early in the pulse, and therefore were insensitive to the
  different peak intensities of the pulses. OBD events that have narrower
  angular distributions ($W_{n\geq2}$)  are found to be markedly dependent on the chirp
  magnitude and sign, with approximately $100\%$ variation.

Dissociation from positively chirped pulses in the range 500-1200 fs$^2$ is found to
be more efficient than from transform limited pulses. Chirped pulses with GDD=1320
fs$^2$ (120 fs duration) produce a nearly equal number of OBD events compared with
transform limited pulses, even though the peak intensity is smaller by a factor of
four. At GDD values higher than 1320 fs$^2$, the number of OBD events diminishes due
to the decreasing peak intensity inherent in exceedingly elongated pulses.  In order
to neutralize the influence of peak intensity on the efficiency of dissociation, one
can examine the ratio of positive to negative chirp dissociation events of the same
chirp magnitude, hence the same peak intensity. At $1600$ fs$^2$, for example, we
find that positively chirped pulses induce over the barrier dissociation up to $55\%$ more efficiently
than negatively chirped pulses, whereas near-resonant dissociation events are barely
affected by the sign of the chirp.

The optimal chirp rate for maximizing the dissociation by bond softening depends on both the pulse fluence and the peak power. As discussed earlier, the increase in OBD signal for positively chirped pulses is attributed to the combined effect of dynamic alignment of molecules as well as of lowering of the barrier. The first part is pulse duration dependent. Three interlinked parameters characterise the laser pulses, namely the peak power, linear chirp (which in turn controls the pulse duration) and pulse fluence. For the current set of experiments, where the third parameter is kept constant, the effect depends on the first two parameters in opposite ways. As the chirp is increased (which in turn increases the pulse duration and reduces the peak power), more alignment is expected despite of the somewhat lowered peak power \cite{esry2009}. However, the gap opening in the potential energy curves reduces with peak power, which makes bond softening less efficient. This is reflected in the drop in the dissociation signal at higher chirp values. It is expected that the optimal chirp for OBD will shift to higher values as the fluence is increased simply because more molecular alignment will be achieved.

\section{Conclusion}

To conclude, we have demonstrated  that strong field photodissociation of H$_2^+$ is
controllable by varying the sign and magnitude of a linearly chirped intense
ultrashort pulse. We suggest a simple mechanism to explain this control, using a dynamically dressed potential curves picture. This approach enhances or suppresses photodissociation where
bound-to-repulsive transitions dominate. Linear chirp is not necessarily
 optimal for controlling dissociation, as higher
orders of dispersion can be used to optimally tailor the
instantaneous frequency and intensity envelope. Future experiments, for example,
could combine third-order dispersion with GDD to yield asymmetric temporal envelopes
with frequency sweeps that may further enhance dissociation.

\begin{acknowledgments}
We thank B. D. Esry, and N. Moiseyev for helpful discussions, and K.D. Carnes for comments on the manuscript. This work was partially
supported by the Israeli Science Foundation.  IB and BDE are thankful for support by the Chemical Sciences, Geosciences,
and Biosciences Division, Office of Basic Energy Sciences, Office of Science, US Department of Energy.
We also acknowledge support for the WIS-JRML for this collaboration by the
U.S.-Israel Binational Science Foundation (BSF). D. Sch. is acknowledging support by the Weizmann Institute of Science  through the Joseph Meyerhoff program.
\end{acknowledgments}

\end{document}